\documentclass{epl}

\title{Information filtering via Iterative Refinement}
\shorttitle{Collaborative filtering}
\author{P. Laureti\inst{1} \and L. Moret\inst{1} \and Y.-C. Zhang\inst{1} \and Y.-K. Yu\inst{2}}

\institute{                    
  \inst{1} D\'epartement de Physique, Universit\'e de Fribourg - CH-1700 Fribourg, Switzerland.\\
  \inst{2} National Center for Biotechnology Information, NIH - 8600 Rockville Pike, Bethesda, MD 20894.
}

\newcommand{\Avg}[1]{\left\langle{#1}\right\rangle}
\newcommand{\avg}[1]{\left\langle{#1}\right\rangle}
\newcommand{\ave}[1]{\langle{#1}\rangle}
\newcommand{\Ave}[1]{\left\langle{#1}\right\rangle}
\newcommand{\be}{\begin{equation}}
\newcommand{\ee}{\end{equation}}
\newcommand{\beas}{\begin{eqnarray*}}
\newcommand{\eeas}{\end{eqnarray*}}
\newcommand{\bea}{\begin{eqnarray}}
\newcommand{\eea}{\end{eqnarray}}
\newcommand{\req}[1]{(\ref{#1})}
\newcommand{\comment}[1]{}
\def\C{\mathcal{C}}
\def\D{\mathcal{D}}

\begin{document}

\maketitle

\begin{abstract}
With the explosive growth of accessible information, expecially on the
Internet, evaluation-based filtering has become a crucial task. Various
systems have been devised aiming to sort through large volumes of
information and select what is likely to be 
more relevant. In this letter we analyse a new ranking method,
where the reputation of information providers is
determined self-consistently. 
\end{abstract}

\section{Introduction}
The study of complex networks and of some dynamical processes taking place
on these structures has recently attracted a great deal of
attention in the physics community~\cite{WS98, AB02, DM02, Newman03d}. 
The importance of technological networks, such as the Internet, lies mostly in the increased
communication capabilities~\cite{RS03, ADG00}, which make information
progressively easier to produce and distribute.
As storage and transmission costs continue to drop,
an overabundance of information threatens to overwhelm its
recipients. 
 It is, therefore, crucial to process
information in order to present a user only the one that answers best
her requests~\cite{V95}. 
\comment{
A good filter of information can give enterprises a competitive
advantage~\cite{P85} and provide consumers with a powerful
leverage to improve market efficiency.
}

An important aspect of information filtering regards {\it scoring systems} in
the World Wide Web~\cite{pagerank, kleinberg}. They collect
evaluations and aggregate them into published scores that
are meaningful to the final user. This embraces many
different instances, ranging from commercial websites, where buyers
evaluate sellers (Ebay, Amazon, etc.) to new generation search
engines (Google, Yahoo, etc.), and
opinion websites, where people evaluate
objects (Epinions, Tailrank, etc.) 
Since the evaluators carry different expertise, 
it is important to estimate how accurate a given vote may be and to
weight it accordingly. This can be done through the use of
raters' reputations~\cite{colfil}. 
{\it Reputation} summarises one's past behaviour and has always been used to
bear the risk of interacting with strangers. 
The Internet, while
enhancing such a risk, brings in the possibility to find its
antidotes~\cite{rep}.
Since nobody knows a-priori who are the honest and competent
evaluators, in fact, online scoring systems often include some measure of
their past performance.
This gives users an
indication on how trustworthy a given piece of information is supposed
to be. An expert of the field would probably obtain a high
reputation; experts' votes should then count more when aggregating the scores.
While reputation is usually obtained by asking users supplementary
evaluations about other users, the procedure of {\it Iterative
Refinement} ({IR}), which can be shown to outperform naive
methods~\cite{yzlm}, does not require to explicitly rate the raters.

 The aim of this letter is to study, in a generalised model,
 the {IR} method's dependence on the relevant parameters,
illustrate the subtle issues in 
 its mathematical underpinning and elaborate on distortions generated
 by different kinds of cheating. Prior to describing 
  the major focus of this work, we will briefly state the model   
   and define some notations. 

\section{Model and algorithm}
 To describe our approach in the simplest manner, let us consider 
 $N$ raters evaluating $M$ objects, which can be books, movies or even
other raters. Each object $l$ has
an intrinsic quality $Q_l$ and each rater $i$ has an intrinsic judging power 
 $1/\sigma_i^2$. Let $x_{il}$ be a random variable representing the
 rating given by rater $i$ to object $l$. Intrinsic qualities and
 judging powers are defined by the first two moments of its distribution:
\bea
\label{avg}
&&\Avg{x_{il}} = \mu_{il} = Q_l+\Delta_{il}\\ \label{var}
&&\Avg{(x_{il}-\mu_{il})^2} = \sigma_i^2 \;,
\eea
where $\Delta_{i l}$ is the systematic error of agent $i$ 
towards object $l$. Expectation values are taken over the distribution of
$x_{il}$. They can be regarded as ensemble averages, obtained if
 the evaluations were to be performed infinitely many times.
Our aim is then to extract the quality of each object from a single set $\{x_{il}\}$
  of evaluations. We thus estimate the intrinsic quality $Q_l$ of 
 object $l$ by a weighted average of the received votes
\be\label{mul} 
 q_l = \sum_{i=1}^N f_i x_{il}\; ;
\ee
the inverse judging power $\sigma_i^2$ of rater $i$ is 
estimated by the sample variance $V_i$ 
\be\label{sigma}
V_i= \frac{1}{M} \sum_{l=1}^M (x_{il}- q_l)^2 \;.
\ee
The unnormalised weights $\omega_i$ take the general form 
\be\label{weights}
\omega_i = {V_i}^{-\beta} \;,
\ee
with  
$\beta \geq 0$ and $f_i=\omega_i / \sum_j \omega_j$. 
As such, $\omega_i$ decreases
  when $V_i$ increases because rater $i$ has a lower
judging power and should be given less credit. We will consider
  scenarios where $\beta$ equals $1$ or $1/2$.
The case $\beta = 1/2$, in fact, exhibits scale-changing and translational
 invariance because $q_l$ becomes a sum of
  dimensionless random variables; the case $\beta = 1$ corresponds to 
 optimal weights, as explained later in the section {\it No systematic
  errors.}

The {IR} algorithm allows 
to solve eqs.~(\ref{mul}-\ref{weights}), thus estimating $Q_l$ and $\sigma_i$, via the following recursive procedure:
{\it I}) Without additional information, set
  $\omega_i=1/N \; \forall i=1,2,...,N$. {\it II}) Estimate $q_l$ with
eq.~\req{mul}. {\it III}) Estimate $V_i$ with eq.~\req{sigma} and plug
it in  eq.~\req{weights} to find the weights. {\it IV})
Repeat from step {\it II}. Numerical simulations show that this
process converges to the minimum of the cost function
${E}(\{q_l\})=\sum_{i} \left[ \sum_{l} (x_{il}-q_l) V_i^{-\beta} \right]^2$ much
faster than other conventional methods.

\section{Analytical approach}
Eq.~(\ref{var}) implies that the random variable   
$(x_{il}-\mu_{il})^2$ has mean $\sigma_i^2$ and 
  variance $m_i^2 \sigma_i^4$, which is determined by 
 the distribution of $x_{il}$; in particular, $m_i^2=3$ if the
 distribution of votes is itself Gaussian. Let us define the variable 
$\gamma_{ij}= \frac{1}{M}\sum_{l} (x_{il}-Q_l) (x_{jl}-Q_l)$;
provided that $x_{il}$ has finite moments of, at least,
order $4$, in the large $M$ limit one obtains
\be \label{central}
\gamma_{ij}
 \to  \sigma_i^2  \delta_{ij} 
 + \overline{\Delta_i \Delta_j} + \frac{1}{\sqrt{M}} (e_{ij} +\overline{\Delta_i} h_j +
 \overline{\Delta_j} h_i) \;.
\ee
Here the overlined quantities represent averages over the $M$ items,
$\overline{x_i}=\frac{1}{M}\sum_l x_{i l}$.  
The Gaussian random variables $e_{ij}$ and $h_i$ have mean zero 
 and variances
${\ab{var}}(e_{ij})= m^2_{ij} \sigma_i^2 \sigma_j^2 $ and
${\ab{var}}(h_{i})= \sigma_i^2$,
where $m^2_{ij}=1+\delta_{ij}(m_i^2-1)$. 
In the following we shall use the notation $g_i=e_{ii}$.

Eq.~\req{central} has to be interpreted in
 probability, as prescribed by the Central Limit Theorem. 
\comment{
Deviations from the
 Gaussian behaviour
 are possible in the tail of the distribution of the random variables,
 but increasingly less probable as $M$ grows. 
}
In its derivation we have further assumed that
 raters are independent; in fact, the correlation among the variables $e_{ij}$ of different
 indices diminishes as $M$ increases. If $M \gg 1$, 
 the random variables $\{ e_{ij} \}$ are effectively 
 independent, with the first visible triangular 
 correlation of order $1/M^2$ or smaller. From counting the 
 degrees of freedom associated with random numbers, 
 it is desirable to have $M \ge (N+1)/2$. 
\comment{ The set $\{ (x_{il}-\mu_{il}) \}$  constitutes $MN$ random numbers, 
 while the total number of $e_{ij}$ is $N(N+1)/2$.}
 Eq.~(\ref{central}) forms the basis of 
 our analytical pursuit in the later development. 

The performance of the {IR} method can be stated by measuring the
following mean squared errors: 
\bea \label{dq}
d_{q_{l}}&=& \avg{(q_l-Q_l)^2} = \Avg{(s_l+\tilde{\Delta}_l)^2}
%= {\sf Var} (q_{l})
%  = \frac{1}{M} \sum_{l=1}^M  \avg{(q_l-Q_l)^2}  
\\
\label{ds}
d_{\sigma_{i}}&=&\avg{(V_i-\sigma_i^2)^2}= {\sf Var}
(V_i) + \ab{Bias}^2 (V_i)\; ,
\eea
with $\ab{Bias}(V_i)\equiv \avg{V_i} - \sigma_i^2$. 
In eq.~\req{dq}
we have separated the systematic error part, making use of the variables
$\tilde{\Delta}_l=\sum_i f_i \Delta_{il}$
and $s_l=\sum_j (y_{jl}-Q_l) f_j$, with
$y_{il}=x_{il}-\Delta_{il}$. Eqs.~(\ref{avg},\ref{var}) guarantee that the
first two
moments of $(y_{il}-Q_l)$ are independent of index $l$, therefore $\avg{s_l^2}
 = \frac{1}{M} \sum_{l=1}^M \avg{s_l^2}$. This permits us to employ
 eq.~\req{central} to obtain
\be\label{sl}
\Avg{s_l^2}= \sum_{i} \sigma_i^2 \Avg{f_i^2}
+ \sum_{i,j} \Avg{f_i f_j \frac{e_{ij} }{\sqrt{M}}}.
\ee
The variable $s_l$ becomes Gaussian in the large $N$ limit, 
as long as the weights $f_j$ are fixed and satisfy the Lindeberg 
 condition~\cite{Feller}.  However, 
such inference can't be drawn easily because the weights 
 and the estimated $q_l$ are tangled up in eqs.~(\ref{mul}-\ref{weights}).  
 The standard deviation of $s_l$ can, nevertheless, 
 be calculated.  
The general problem of finding intrinsic values from
completely distorted votes is not solvable. 
In fact, even if one disposed of an infinite number of raters
and evaluations, the estimator \req{mul} of $Q_l$ would always be
biased of the amount $\langle\tilde{\Delta}_l\rangle$.
We shall, in the following, focus our attention on three particular cases of special interest.

\section{No systematic errors}
When $\Delta_{il}=0 \; \forall i,l$,
raters are impartial but posses different judging powers. 
In order to obtain the best quality estimator one can minimise
the mean squared error 
$d_q(\{\omega_k\})$ of \req{mul} with respect to the $\omega_i$'s.
This gives the optimal weights~\cite{Hoel}, $\beta = 1$ in (\ref{weights}), 
  with minimal
$d_q(\{1/\sigma^2_k\})= 1/\sum_i \sigma_i^{-2}$.
\comment{ 
If one plugs in the objective function $d_q(\{\omega_k\})$ the weights given by  
$\beta = 1/2$, the resulting error will be larger. 
To be specific, $d_q(\{1/\sigma_k\}) = N/ 
\left(\sum_i \frac{1}{\sigma_i}\right)^2 \to
\frac{1}{N} \Avg{\frac{1}{\sigma}}_d^{-2} \ge d_q(\{1/\sigma^2_k\})$.
--and the same is true for their fluctuations. 
}
 Since the law of large numbers guarantees the convergence of 
  $d_q(\{1/\sigma_k\})$ to zero for large $N$, the same must obviously be
  true for optimal weights. 
\begin{figure}
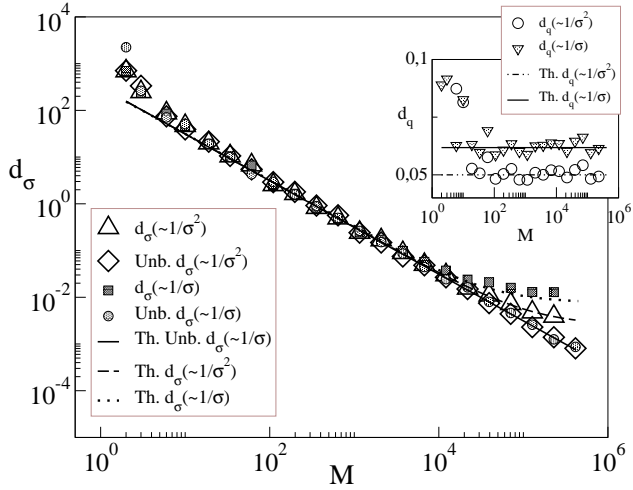
%[ht!]
%\centerline{\includegraphics{dsvsminPaper.eps}}
\onefigure[width=8.3cm]{dsvsminPaper.eps}
\caption{Average squared difference $d_{\sigma}$ between given and predicted
variance, as a function of $M$ in log-log scale. 
Symbols represent simulations of the IR method with  $\beta=1$
(triangles) and
$\beta=1/2$ (filled squares) in eq.~\req{weights}; 
Diamonds and filled circles show simulations of $d_{\sigma}$, where the
  estimator of the variance has been corrected for the
  bias. The corresponding theoretical predictions, calculated as
  explained in the text, fit the data very well.
In the inset a similar plot shows the coincidence between the
predicted and simulated plateau reached by $d_q$ for large $M$.
Parameters of the simulations: $N=100$, 
 intrinsic values $Q_l$ uniformly
  distributed between $10$ and $20$ and standard deviations $\sigma_i$ uniformly 
distributed between $1$ and $5$; averaged over $10^3$ realizations.}
%\vspace{-0.5in}
\label{fig2}
\end{figure}
Unfortunately, it is not possible to state that the choice 
$\beta=1$ is optimal if  
the $\sigma_i^{2}$'s  are not known in advance.
Although the convergence of $q_l \to Q_l$ for $N \to \infty$
 is guaranteed, % by the law of large numbers, 
 the small deviation $|q_l - Q_l|$ 
  due to finite $N$ will propagate to the estimate of
 $\sigma_i^2$ and render $V_i \ne \sigma_i^2$, even when 
 $M \to \infty$. 
A recursive procedure allows to calculate the expectation values for $\avg{f_i}$;
using eq.~\req{central}, it is straightforward to show that 
\be
V_i %&=& {1\over M} \sum_{l=1}^M (x_{il}-q_l)^2 
%= {1\over M} \sum_{l=1}^M \left[ (x_{il}-Q_l) + (Q_l-q_l) \right]^2 
%\nonumber \\
= \left[ \sigma_i^2 + \frac{g_i}{\sqrt{M}} \right] \left( 
      1- 2 f_i \right) + 
      \sum_j f_j^2 \left( \sigma_j^2 + \frac{g_j}{\sqrt{M}} \right) 
     \;\; + 2 \sum_{j<k} f_j f_k %{\omega_j  \omega_k \over (\sum_n \omega_n^2)^2}
	  \frac{e_{jk}}{\sqrt{M}} 
            - 2 \sum_{j;j \ne i} f_j %{\omega_j \over \sum_k \omega_k}
\frac{e_{ij}}{\sqrt{M}}.
    \label{vi}
\ee
Now we use $\omega_i = V_i^{-\beta}$ and, after iterative substitutions, 
 we may express $\omega_i$ in terms of $\sigma_i$'s and random
 variables $\{e_{ij}\}$. 
One may then compute $f_i$ and plug it in eqs.~(\ref{dq}-\ref{sl}).
Let us define
$G(b) \equiv \frac{1}{N} \sum_i m_i^2 \sigma_i^{-b}$ and 
 denote by angular brackets a simple average over
the raters $\ave{y}=\frac{1}{N}\sum_i y_i$. 
 Equipped with this formalism, we perform tedious
 but straightforward calculations to obtain
 the following asymptotic expansions, for $M,N \to \infty$, to
 the first two dominating orders:
\bea
&&\Avg{(q -Q)^2} \simeq \frac{1}{N \ave{ \sigma^{-2\beta}}^2}
 \left[\Ave{\frac{1}{\sigma^{4\beta-2}}}  + \frac{\beta \C_1}{N} +
   \frac{\beta \C_2}{M}\right] 
 ,\label{dq_final} 
\\
&&\ab{Bias} (V_i) \simeq \Avg{(q-Q)^2} 
- \frac{2\sigma_i^{2-2\beta} }{ N \ave{\sigma^{-2\beta}}}
 \left[ 1 + \frac{\beta \D_1 }{ N} + \frac{\beta \D_2 }{ M} \right]
,\label{bias}  \\
&&{\sf Var}(V_i) \simeq m_i^2 \frac{\sigma_i^4 }{ M}
 \left[ 1 + \frac{(\beta +1) \sigma_i^{-2\beta} 
 }{ N \ave{\sigma^{-2\beta}}} \right] 
,\label{ds_final}   
\eea
with complicated constant coefficients~\footnote{ 
They are given by: 
 $
\C_1=
 \frac{4 \ave{\sigma^{2-6\beta}}}{\ave{\sigma^{-2\beta}}} + 
 2 \frac{\ave{\sigma^{2-4\beta}}^2 \ave{\sigma^{-2(\beta+1)}}}{\ave{\sigma^{-2\beta}}^3} %\right. \\ \left.
-  6 \frac{\ave{\sigma^{-4\beta}} \ave{\sigma^{2-4\beta}}}{
   \ave{\sigma^{-2\beta}}^2} $, 
$
\C_2 = 4 \langle\sigma^{2-4\beta}\rangle 
 + (2\beta -1) G(4\beta-2)  
-  \frac{(\beta+1) G(2\beta) \ave{\sigma^{2-4\beta}} }{
\ave{\sigma^{-2\beta}}}
$, 
$
\D_1 = \frac{2 \sigma_i^{-2\beta} 
}{\ave{{\sigma^{-2\beta}}}}
- \frac{\ave{{\sigma^{2-4\beta}}} \sigma_i^{-2}}{
 \ave{{\sigma^{-2\beta}}}^2}
  - \frac{2 \ave{\sigma^{-4\beta}}}{\ave{{\sigma^{-2\beta}}}^2}
+
 \frac{\ave{{\sigma^{2-4\beta}}} \ave{{\sigma^{-2(\beta
	 +1)}}}}{\ave{{\sigma^{-2\beta}}}^3}
$ and 
$
\D_2= \frac{(\beta -1)m_i^2}{2}
  - \frac{(\beta +1) G(2\beta)}{\ave{{\sigma^{-2\beta}}}} 
 + 2\frac{\ave{{\sigma^{-2(\beta+2)}}} \sigma_i^{-2\beta}}{\ave{\sigma^{-2\beta}}}
$. }.
These expressions simplify  
 considerably when taking the limit $\beta = 1/2$ and $\beta =
 1$. For instance, eq.~\req{bias} takes the forms  $\ab{Bias}_{\beta=1}(V_i)
\simeq  - 1 / \left( N \Ave{\sigma^{-2}}\right) \label{bias1LL}$
 and $\ab{Bias}_{\beta=1/2}(V_i) \simeq 1/(N\ave{1/\sigma}^{2}) 
- 2\sigma_i / \left( N \ave{\sigma^{-1}}\right)$.
The analytical solution allows one to find an {\it unbiased  estimator} 
for $\sigma_i^2$ --up to ${\cal O}(1/N^2, 1/NM)$. 
In applications we  may use eq.~\req{sigma} as an estimator of $\sigma^2$
 to evaluate $\ab{Bias}(V_i)$ and  redefine the weights as 
%estimate the bias, using eq.~\req{sigma} as 
% an estimator of $\sigma^{2}$, and redefine the weights as
$\omega_i=1/(V_i-\ab{Bias}(V_i))$. 
Since we have here $d_{q_l}=s_l$, suffices
 to plug eqs.~(\ref{dq_final} - \ref{ds_final}) in eq.~\req{sl}  
to find theoretical expressions for the mean squared errors.
They are shown to match 
numerical simulations in figs. \ref{fig2} and \ref{fig1}.

In fig.~\ref{fig2}, the mean squared error of the variance
$d_{\sigma}=\frac{1}{N}\sum_i d_{\sigma_i}$ is plotted against
$M$ in log-log scale. 
Our theoretical prediction becomes very good as soon as $M>10$.
Diamonds and filled circles show simulation
 results of the {IR}
method where the biased estimator of the variance has been corrected by
recursive use of eq.~\req{bias}:
the plateau reached by $d_{\sigma}$ for large $M$ disappears because
the accuracy of the prediction can be thus improved by two orders of
magnitude. The mean
squared error of the quality $d_{q}=\frac{1}{M}\sum_l
d_{q_l}$, on the other hand, can never vanish for large $M$ when $N$ is finite. This is
shown in the inset of fig.~\ref{fig2}, while the dependence of $d_{q}$ on $N$ is
reported in fig.~\ref{fig1}. We have also plotted therein, as a dotted
line, the behaviour of the same quantity when the estimator of $q_l$ is just the average
unweighted vote received by item $l$. This illustrates how IR is able to
reduce the error.
A comparison between  the two weighting schemes 
shows that $\omega_i = 1/V_i$ performs almost always better
than $\omega_i = 1/\sqrt{V_i}$. The inset of fig.~\ref{fig1} shows
$d_\sigma$ vs. $N$; the
plateau, which is the same for $\beta=1/2$ and $1$, vanishes
for $M\to \infty$ when corrected for the bias as before.
\comment{The only exception
appears for very small $M$ values, as in the upper left corner
of fig. \ref{fig2}. This feature, not predictable by asymptotic
analytical estimations, disappears if the distribution of
$1/\sigma_i$ has fat tails. Broader distributions of raters' judging
powers actually favour the $\beta=1$ scheme progressively.}
\begin{figure}
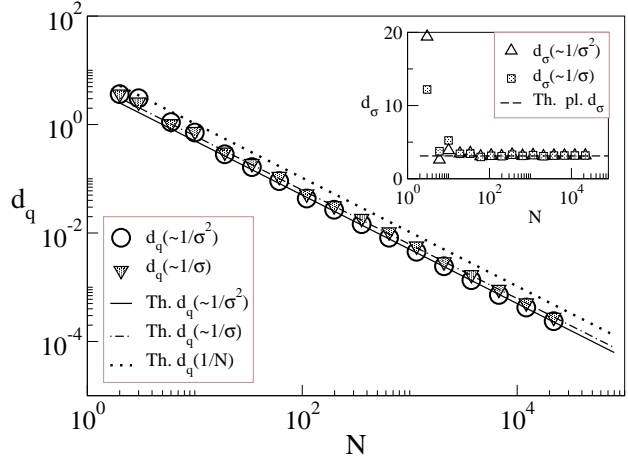
%[ht!]
\onefigure[width=8.2cm]{dqvsnin.eps}
\caption{Average squared difference between estimators and intrinsic values,
for  quality (main) and variance (inset),
plotted in log-log scale as a function
of $N$, with $M=100$, for $\beta=1/2,1$. 
Symbols represent simulation results of the IR method,
%with intrinsic values distributed as in fig.~\ref{fig2}, 
lines are the corresponding theoretical predictions. The dotted line
represents $d_q$ when the quality estimator is just the straight
average.}
\label{fig1}
\end{figure}

\section{Camouflage}
Let us now restart from the general problem of
eqs.~(\ref{avg},\ref{var}). The case we want to analyse here is that
of ratings affected by systematic
errors that depend on the rater but not on the ratee, $\Delta_{il}=\Delta_i \; \forall l$. 
Such a fictitious distortion
is instructive to study analytically and can be easily generalised to more
interesting cases. In fact, as it alters a rater's scale of evaluation
but not the ranking of her preferences, it can serve as a basis to
study systems where agents are only asked to sort a set of items in order
of increasing quality.

If one knew the values of $\Delta_i$ for all $i$, one could find
 the optimal weights $\{\omega_k^*\}$ proceeding
 as described in absence of systematic errors. 
 Upon minimisation of
$d_q(\{\omega_k\})$ with respect to the
$\omega_i$'s one obtains
%\bea\label{opt}
$\underline{\omega}^* = A^{-1} \underline{1}$, with
$A_{ij} = \sigma_i^2 \delta_{ij} + \Delta_i \Delta_j$. Here we have
 used a more compact matrix notation, where
$\underline{1}$ is a vector of ones.
\comment{
The mean squared error then reads
%\be\label{mse}
$\min_{\omega} d_q(\{\omega_k\})= 1/\sum_{ij} (A^{-1})_{ij}$.
}

Whenever the deviations $\Delta_i$ are
small, limited to a minority of the population or randomly
distributed around zero, they can be somehow detected. In the general
case  one can only detect, at best, the {\em relative} systematic errors.
In fact %unlike the case of eq.~\req{dq}, 
$\tilde{\Delta}=\sum_j f_j \Delta_j$ does not depend
on $l$ in presence of camouflage and
the relevant quantities only depend on $\Delta_i$ under the form 
$\delta_i=\Delta_i-\tilde{\Delta}$. For instance,
the variance can be written as 
$
V_i=\frac{1}{M} \sum_l \left[ \sum_j f_j ( y_{jl} - y_{il} ) +
  \delta_i \right]^2
$.
This means that, if we change the $\Delta_i$'s while keeping the
$\delta_i$'s unchanged, we end up with the same result for $d_q$,
only translated by the amount $\tilde{\Delta}$. 

In order to estimate analytically the performance of the {IR} method,
we can posit $\tilde{\Delta}=0$
and solve eqs.~(\ref{mul}-\ref{weights}) as before. Thus we find
$f_i(\{\delta_i\})$, whose term of order zero is $(\sigma_i^2+\delta_i^2)^{-\beta} / \sum_j 
(\sigma_j^2 + \delta_j^2)^{-\beta} $.  
This way we have a formal solution as a function of
$\delta_i$, which must comply with the
constraint $\sum_i f_i \delta_i=0$ and can eventually be recovered numerically.
\comment{
We
have successfully tested a numerical
procedure to recover the $\delta_i$-s, which consists in iterating equation
$\delta_i^{(n)}=\Delta_i-\sum_j f_j^{(n-1)} \Delta_j$,
with $f_j^{(n-1)} = f_i(\{\delta_i^{(n-1)}\})$ for $n>1$ and
$\delta_i^{(0)} = 1/N$.
}
\begin{figure}
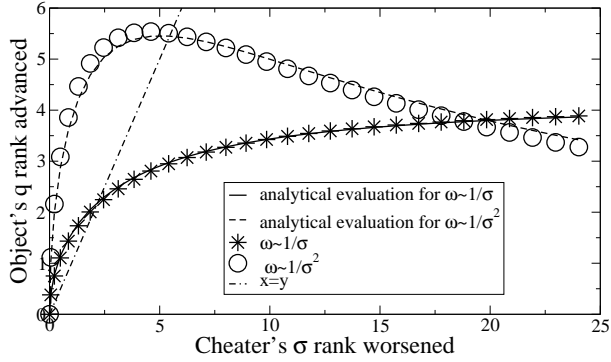
%[ht!]
\onefigure[width=8cm]{biasFixedSigmaNoErrorBars.eps}
\caption{Increase of object's quality as a function of the cheater's rank
 loss, as the value of $\Delta$ grows from $0$ to $30$. 
Simulations have been carried on with 
$N=100$, $M=100$ and intrinsic values distributed
as in fig.~\ref{fig2}, except for $\sigma_I = 1$ and $Q_L=20$.
The theoretical estimations are parametric plots of
 eq.~\req{biasEq} for $\beta=1$ and $1/2$.}
\label{fig3}
\end{figure}

\section{Cheating}
It is interesting to consider the case of one intentional cheater $I$
wanting to boost the value of object $L$ of an amount
$\Delta$, all other raters being honest: $\Delta_{il}= \delta_{iI}\delta_{lL}\Delta$. 
Agent $I$ commits no systematic error in evaluating all objects
but $L$. Still, she would loose credibility and weight as 
$\Delta$ becomes larger; this would eventually diminish her relative
influence over object $L$. It is important to evaluate the 
difference $\delta q_l=q_l(\Delta) - q_l$  between the estimated 
value of the object with and without
the friendly uprating. In fact a small $\delta q$, compared to the
lost in credibility of the rater, discourages cheating, and
vice-versa. 

The variance, as defined in eq.~\req{sigma}, can be written as a
function of $\Delta$ and of the normalised weights. Hence
$
V_i(\Delta,\{f_i(\Delta)\})=V_i(0,\{f_i(\Delta)\})+ \delta_{iI} \Delta^2/M
$, 
where the formal expression of $V_i(0,\{f_i(\Delta)\})$ is equal to
that of eq.~\req{vi}. 
Iterative asymptotic expansions can be
performed the same way we did in absence of systematic errors. 
In this case the variables $y_{il}$
are equal to the $x_{il}$, except for $y_{IL}=x_{IL}-\Delta$. 
Therefore eq.~\req{mul} becomes
$q_l(\Delta) = \sum_i f_i(\Delta) y_{il} + \Delta \delta_{lL}
f_I(\Delta)$, which implies
$q_l(\Delta) - q_l \simeq \Delta \delta_{lL} f_I$. 
For $\Delta \ll \sqrt{M}$ the average
deviation reads
$
\Avg{\delta q_L}  \simeq \Delta 
\left( 
\avg{f_I} - \beta \frac{ \Delta^2}{\sigma^2_I M}
\right)$.
%,where $f_i\equiv f_i(\Delta=0)$.
If the value of $\Delta$ is comparable to $\sqrt{M}$, on the other
hand, the zeroth order of the correction at the thermodynamic limit amounts to
\be\label{biasEq}
\Avg{\delta q_L} \to \Delta \cdot \left[ 
\left( \sigma_I^2 + \frac{\Delta^2}{M}
  \right)^{\beta}\, \sum_{j \ne I} \sigma_j^{-2\beta} +1 
\right]^{-1}.
\ee
\comment{
As a consequence, the maximum result is not
obtained by attributing the
highest possible vote to $L$, but by setting 
$\Delta=\sigma_i \sqrt{M}$. 
}
In fig.~\ref{fig3} eq.~\req{biasEq} is shown to fit the simulations
fairly well in the rank space. We have compared thereby the scheme $\beta=1$ (circles) with
$\beta=1/2$ (stars) in the worst case: the best agent
is trying to raise the worst object. 
In the region of moderate
cheating the $\omega_i = 1/\sqrt{V_i}$ weighting scheme is less
sensitive to cheating. This is particularly important left to the
$x=y$ line, where the cheater pays less then what she offers to
the object and cheating can be advantageous. 
However, the relative influence of the cheater is a growing, although saturating,
function of $\Delta$. Under the $\omega_i = 1/{V_i}$  weighting
scheme, on the other hand, such an influence starts decreasing once
passed a crossover value. There the cheater's reputation is
so much damaged by her misbehaviour that, if she attributed  a higher
value to object $L$, its estimated rank would diminish. Optimal
weights are, therefore, much more resilient to severe cheating.
%
%This transition can be analytically evaluated
%$\Delta_c=\sqrt{M} \sqrt{ \left[ \Ave{1/\sigma}/\Ave{1/\sigma^2}
%\right]^2-\sigma_{I}^{2}}$
%

We just remark that, taking averages without refinement, a
 cheater would indefinitely increase an object's rank without
 undergoing any punishment.
%in this respect, the situation is critical. 
The transition to the cheater's unfavorable region 
is the solution of $d_{r(q)}=d_{r_{(\sigma)}}$  in the $\Delta$ space.
\comment{
which in the $\omega =\frac{1}{\sigma^2}$ case gives
\be
\frac{\left(\sqrt{\sigma_{b}^{2}+\frac{B^2}{M}}-\sigma_{b}\right)M}
{l_{\sigma}}=\frac{NB}{\sum _i \frac{1}{\sigma_{i}^{2}} 
\left(\sigma_{b}^2+\frac{B^2}{M}\right)l_q}
\ee
 One gets similar results for $\omega=\frac{1}{\sigma}$.
}
\section {Conclusion}
In this letter we have analyzed a novel scoring system that aggregates
the evaluations of $N$ agents over $M$ objects
by use of reputation and weighted averages.
Agents, as a result, are ranked according to their judging
capability and objects according to their quality. The method can be
implemented via an iterative algorithm, where the intrinsic bias of the
estimators of the weights can be corrected.
We show, with simulations and
analytical results, that the method is effective and robust against
abuses. 
The larger the system, the better is the filtering precision. This
method can be applied in web-related reputation and scoring systems.

\acknowledgments
We thank the reviewers for useful remarks. This work was partially supported by the Swiss National Science
Foundation, through project number 2051-67733,
and by the Intramural Research Program of
 the National Library of Medicine at 
NIH/DHHS.

\end{document}